\documentclass[aps,prl,twocolumn,superscriptaddress]{revtex4-2}
\usepackage[english]{babel}
\usepackage{amsmath,amssymb,braket,pifont,amsthm,lipsum}
\usepackage{amsfonts}
\usepackage[pdfborder={0 0 0}, colorlinks=true, urlcolor=blue,linkcolor=blue, citecolor=blue]{hyperref}
\usepackage{color}
\usepackage{graphicx}
\usepackage{epstopdf}
\usepackage{float}
\usepackage{subfigure}
\usepackage{ulem}
\usepackage{url}
\usepackage{lineno}

\makeatletter

\definecolor{color1}{rgb}{0.0,0,1.0}

\usepackage{babel}
\makeatother

\begin{document}

\title{
Engineering frustrated Rydberg spin models by graphical Floquet modulation
}
\author{Mingsheng Tian}
\affiliation{Department of Physics, The Pennsylvania State University, University Park, Pennsylvania, 16802, USA}
\author{Rhine Samajdar}
\email{rhine\_samajdar@princeton.edu}
\affiliation{Department of Physics, Princeton University, Princeton, NJ 08544, USA}
\affiliation{Princeton Center for Theoretical Science, Princeton University, Princeton, NJ 08544, USA}
\author{Bryce Gadway}
\email{bgadway@psu.edu}
\affiliation{Department of Physics, The Pennsylvania State University, University Park, Pennsylvania, 16802, USA}

\begin{abstract}
Arrays of Rydberg atoms interacting via dipole-dipole interactions offer a powerful platform for probing quantum many-body physics. However, these intrinsic interactions also determine and constrain the models---and parameter regimes thereof---for quantum simulation. Here, we propose a systematic framework to engineer arbitrary desired long-range interactions in Rydberg-atom lattices, enabling the realization of fully tunable $J_1$-$J_2$-$J_3$ Heisenberg models. Using site-resolved periodic modulation of Rydberg states, we develop an experimentally feasible protocol to precisely control the interaction ratios $J_2/J_1$ and $J_3/J_1$ in a kagome lattice. This control can increase the effective range of interactions and drive transitions between competing spin-ordered and spin liquid phases. To generalize this approach beyond the kagome lattice, we reformulate the design of modulation patterns through a graph-theoretic approach, demonstrating the universality of our method across all 11 planar Archimedean lattices. Our strategy overcomes the inherent constraints of power-law-decaying dipolar interactions, providing a versatile toolbox for exploring frustrated magnetism, emergent topological phases, and quantum correlations in systems with long-range interactions.
\end{abstract}

\maketitle

\textit{Introduction.}---In recent years, a variety of programmable quantum simulators have emerged as powerful tools for probing the physics of strongly interacting quantum systems~\cite{georgescu2014,altman2021quantum}. Rydberg atom arrays offer one such promising platform due to their combination of tunable geometries and strong, controllable interactions~\cite{adams2019-rydbergreview}. Rapid advances in experimental techniques have allowed for the precise trapping and manipulation of neutral atoms in arrays, enabling the exploration of a wide range of many-body phenomena, ranging from  correlated quantum phases~\cite{deleseleuc2019-antoine,ebadi2021-lukin,chen2023a-antoine} to nonequilibrium quantum dynamics~\cite{bernien2017-dynamics,bluvstein2021b-dynamics,bornet2023a-dynamics,tom1014-dynamics}. A landmark achievement in this regard was the
preparation
of a topological $\mathbb{Z}_2$ quantum spin liquid (QSL)
state~\cite{semeghini2021-lukin,Samajdar.2021,Verresen.2020}, a highly entangled phase of matter which had been sought after in solid-state materials for nearly half a century~\cite{anderson1973resonating,fazekas1974ground}.

Despite this remarkable progress, an apparent limitation of Rydberg simulators lies in the precise nature of the interactions, which typically decay as a power law with distance. This decay restricts the ability to fully investigate models requiring flexible and tunable long-range interactions~\cite{defenu2023-longrange,patrick2017-topology,viyuela2018-topology,jones2023-topology,sky2023,tian23pr}. 
When appropriately chosen, such interactions can introduce geometric frustration and enhance quantum fluctuations, leading to competing orders and exotic quantum phases in frustrated spin systems~\cite{diep2020frustrated}. In particular, the ability to tune the relative strengths of nearest-neighbor ($J_1$), next-nearest-neighbor ($J_2$), and third-neighbor ($J_3$) interactions, which are highly constrained in real materials, is predicted to enable the exploration of quantum phase transitions among gapped and gapless spin liquids, valence bond solids, and a variety of magnetically ordered states~\cite{sandvik2010,Sheng-kagome-prb,zhu2015a-kagome,he2015,he2015a,zhu2019,he2014a,gong2014-kagome,Yao2018,bintz2024diracspinliquidquantum}.
However, the full control over $J_1$-$J_2$-$J_3$ in spin systems remains challenging due to the difficulty of tuning interaction strengths and ranges while maintaining the necessary levels of coherence and control.

\begin{figure}
    \centering
    \includegraphics[width=\linewidth]{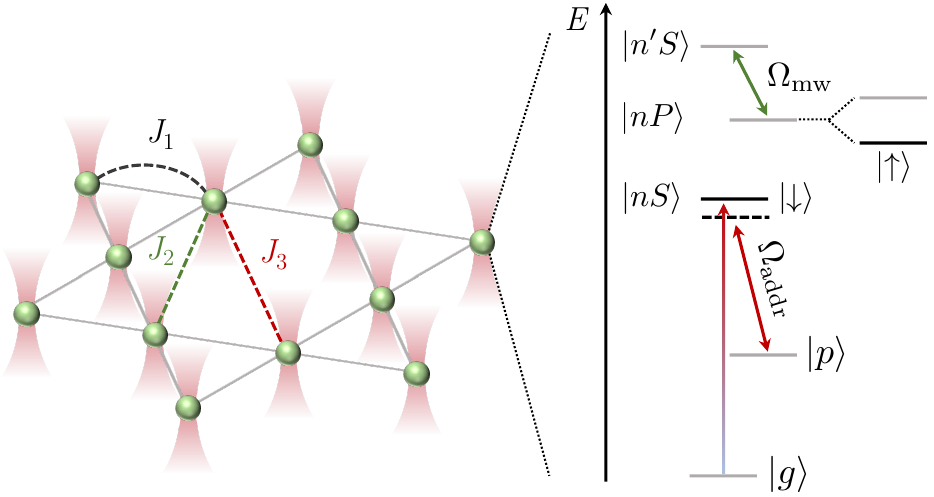}
    \caption{Schematic illustration of $ J_1 $-$ J_2 $-$ J_3 $ dipolar interactions between Rydberg atoms arrayed in programmable geometries, and the corresponding atomic levels relevant for controlling the interaction strengths. 
    The two spin-$1/2$ states are encoded in Rydberg levels, with $\ket{\downarrow} = \ket{nS}$ and $\ket{\uparrow} = \cos\theta\ket{n'S}- \sin\theta\ket{nP}$. The microwave-dressed superposition state $\ket{\uparrow}$ is used to realize the XXZ model, achieved by coupling $\ket{nP}$ to $\ket{n'S}$ using a microwave field with Rabi frequency $\Omega_{\text{mw}}$ and detuning $D$ (with $2\theta = \textrm{arctan}(2\Omega_{\text{mw}}/D)$).
    Addressing beams couple the Rydberg state $\ket{\downarrow}$ to an intermediate state $\ket{p}$ off-resonantly, inducing a local energy shift (dashed line). In the limit of  large detuning $\Delta_{\text{addr}} \gg \Omega_{\text{addr}}$, this shift goes as $\delta \approx  \Omega_{\text{addr}}^2 / \Delta_{\text{addr}}$.
    Precise control of the addressing beams allows for the design of desired $ J_1 $-$ J_2 $-$ J_3 $ interactions.}
    \label{fig1}
\end{figure}

In this work, we develop a systematic framework for engineering long-range interactions in dipolar Rydberg 
arrays,
enabling the realization of the $ J_1 $-$ J_2 $-$ J_3 $ Heisenberg model with precise tunability. By using spatially patterned Floquet engineering of the Rydberg level energies and microwave dressing Rydberg states,
we map the periodically driven system onto an effective undriven system with renormalized interaction strengths, yielding precise control over the transverse and longitudinal spin-spin interactions in the XXZ model.
As a detailed example, we demonstrate how two distinct modulation frequencies can be used to independently control the ratios $ J_2/J_1 $ and $ J_3/J_1 $ on a kagome lattice. We further generalize this method to a wide range of lattice geometries using graph theory, where site-dependent modulations correspond to graph coloring strategies.  
To illustrate its universality, we apply this method to all planar Archimedean lattices, achieving flexible tuning of long-range interactions and overcoming the inherent limitations of power-law decaying interactions. This approach opens a pathway to explore tunable frustrated spin systems and exotic quantum phases induced by long-range interactions in Rydberg arrays.

\textit{System and spin models.}---We begin our discussion with the general spin-1/2 antiferromagnetic $J_1$-$J_2$-$J_3$ Heisenberg model, which is a fundamental model of quantum magnetism. This is described by the Hamiltonian  
\begin{alignat}{3}
H = J^{}_1\hspace*{-0.2cm} \sum_{\langle m, n \rangle} \mathbf{S}^{}_m \cdot \mathbf{S}^{}_n 
+ J^{}_2\hspace*{-0.2cm} \sum_{\langle\langle m, n \rangle\rangle} \mathbf{S}^{}_m \cdot \mathbf{S}^{}_n 
+J^{}_3\hspace*{-0.2cm} \hspace*{-0.2cm}\sum_{\langle\langle\langle m, n \rangle\rangle\rangle}\hspace*{-0.2cm} \mathbf{S}^{}_m \cdot \mathbf{S}^{}_n ,     
\end{alignat}
for $J_1, J_2, J_3 > 0$, 
where $\langle m, n \rangle$, $\langle\langle m, n \rangle\rangle$, and $\langle\langle\langle m, n \rangle\rangle\rangle$ denote nearest-neighbor (NN), second-nearest-neighbor (2NN), and third-nearest-neighbor (3NN) interactions, respectively. The competition between such
interactions introduces geometric frustration and is known to be crucial for the realization of exotic ordered and QSL phases~\cite{Savary-2017, broholm2020quantum}.

To realize and manipulate this model, we consider a system of Rydberg atoms arranged on a lattice, as illustrated in Fig.~\ref{fig1}.
The spin-up and spin-down states are defined by the Rydberg states, with 
$\ket{\downarrow} \equiv \ket{nS}$ and 
$\ket{\uparrow} \equiv \cos\theta\ket{n'S}- \sin\theta\ket{nP}$~\cite{supplementary}, 
where $\ket{\uparrow}$ is encoded as one state from the Autler-Townes doublet 
formed by strongly driving the $\ket{nP} \leftrightarrow \ket{n'S}$ transition with 
Rabi frequency $\Omega_\textrm{mw}$ and detuning $D$ ($\theta=\arctan(2\Omega_{\text{mw}}/D)/2$). 
Using these ingredients, one can simulate a generalization of the Heisenberg model called the XXZ model, which is given by
\begin{alignat}{1}
\nonumber
H^{}_{\textsc{xxz}} = \sum_{\mu} &  \sum_{( m, n ) \in\, \mu} J_\mu^{\pm} ( S_m^+ S_n^- + S_m^- S_n^+ ) + J_\mu^z S_m^z S_n^z \ ,    
\end{alignat}
where $\mu$ indicates the NN, 2NN, or 3NN nature of the couplings. 

In the spin basis, the strengths of the ``flip-flop'' ($J^{\pm}$) and Ising ($J^z$) spin-spin interactions in the XXZ model depend on the exchange terms 
$\sin^2\theta\langle nS, nP | \hat{V}_{\text{ddi}} | nP, nS \rangle$ and $\sin^2\theta\cos^2\theta\langle nP, n'S | \hat{V}_{\text{ddi}} | n'S, nP\rangle$, respectively, with $\hat{V}_{\text{ddi}}$ representing the dipole-dipole interaction operator; see details in the Supplemental Material (SM)~\cite{supplementary}. 
A flexible choice of these matrix elements and microwave-dressed states allows us to globally tune the relative strengths of transverse and longitudinal interaction terms~\cite{supplementary,kurdak2025-dressed,young2021-dressed} in this dipolar XXZ model.
Importantly however, due to the $1/R^3$ power-law decay of the dipolar interactions with distance ($R$), the coupling strengths $J_2$ and $J_3$ are typically much smaller than $J_1$ and the interaction ratios are fixed. This limits the ability to explore the full parameter space of the $J_1$-$J_2$-$J_3$ Heisenberg model and access the rich, frustration-induced physics facilitated by long-range interactions in the first place.

\textit{Floquet engineering}.---To implement the flexible control of long-range interactions,  we use periodically time-dependent local modulations to engineer an effective Hamiltonian. 
This approach is analogous to the shaking of optical lattices, which has been used to realize synthetic gauge fields~\cite{aidelsburger2011-flouqet,struck2012-flouqet,aidelsburger2013,aidelsburger2015-flouqet,eckardt2017} and for band-structure engineering~\cite{sedrakyan2015a-floquet,bracamontes2022-floquet,nixon2024individually}. In Rydberg tweezer arrays, it has previously been used to control many-body dynamics~\cite{bluvstein2021b-dynamics} and extend the Rydberg blockade range~\cite{Zhao2023}. 
As shown in Fig.~\ref{fig1}, addressing light (red) is used to locally shift the energy of the spin-down state.
This is realized by an AC Stark shift, i.e., by off-resonantly coupling to an intermediate state $|p\rangle$ (e.g., the intermediate state of a two-photon ground-Rydberg transition, as in Ref.~\onlinecite{shift-sylvain}). For a large detuning $\Delta_{\text{addr}} \gg \Omega_{\text{addr}}$ (such that induced decay~\cite{supplementary} can be ignored), the $|$$\downarrow\rangle$ state experiences an energy shift of $\delta = \Omega_{\text{addr}}^2 / \Delta_{\text{addr}}$. By modulating the Rabi frequency $\Omega_{\text{addr}}(t)$ using fast acousto- or electro-optics, one creates an effective target Hamiltonian with renormalized interaction strengths.

Specifically, 
we consider a periodic onsite energy modulation on the Rydberg state $|nS\rangle$, as described by  
\begin{equation}\label{eqv}
V(t) = \sum_{m,q} \delta_q \sin(\omega_q t + \phi_{mq}) |nS\rangle \langle nS|,
\end{equation}
where
$m$ indexes the lattice sites and $q$ indexes the modulation frequencies.
Here, $\delta_q$ is the modulation amplitude, $\phi_{mq}$ is the corresponding phase at site $m$, and
$\omega_q$ is the modulation frequency.
The time-dependent modulation of the energy difference between state configurations such as $\ket{\downarrow}_m \ket{\uparrow}_n$ and $\ket{\uparrow}_m \ket{\downarrow}_n$ leads to a renormalization of the interaction between spins on sites $m$ and $n$, by effectively diluting their exchange coupling via the transfer of weight to the nonresonant components of Floquet-dressed states (in contrast to the typical use of Floquet dressing to activate otherwise nonresonant interaction terms~\cite{Clark2019,Miyake,aidelsburger2011-flouqet}).
One can show that the driven system is well-approximated by an effective time-independent Hamiltonian obtained via a unitary transformation~\cite{eckardt2005,lignier2007},
with spin-exchange strengths rescaled as
\begin{equation}\label{eqj}
     J_{mn}^{\text{eff}} = J^{}_{mn} \prod_q \mathcal{J}^{}_0 \left( \frac{2\delta_q}{\omega_q} \sin\left(\frac{\phi_{nq} - \phi_{mq}}{2}\right) \right),
\end{equation}
where $J_{mn}$ denotes the original coupling without modulation,
$\mathcal{J}_0$ is the zeroth-order Bessel function, and $\phi_{mq} - \phi_{nq}$ represents the phase difference between sites $m$ and $n$ for the frequency $\omega_q$ (with frequencies set to be incommensurate in the case of multiple frequencies~\cite{supplementary}).
This enables independent tuning of the transverse spin-spin interaction terms. Such flexible control allows us to explore exotic, otherwise inaccessible quantum phases in many-body systems, including the distinct spin-liquid phases of the kagome spin-1/2  antiferromagnetic XXZ  model~\cite{he2015,zhu2015a-kagome,he2015a}.

\begin{figure}  
    \centering  
    \includegraphics[width=\linewidth]{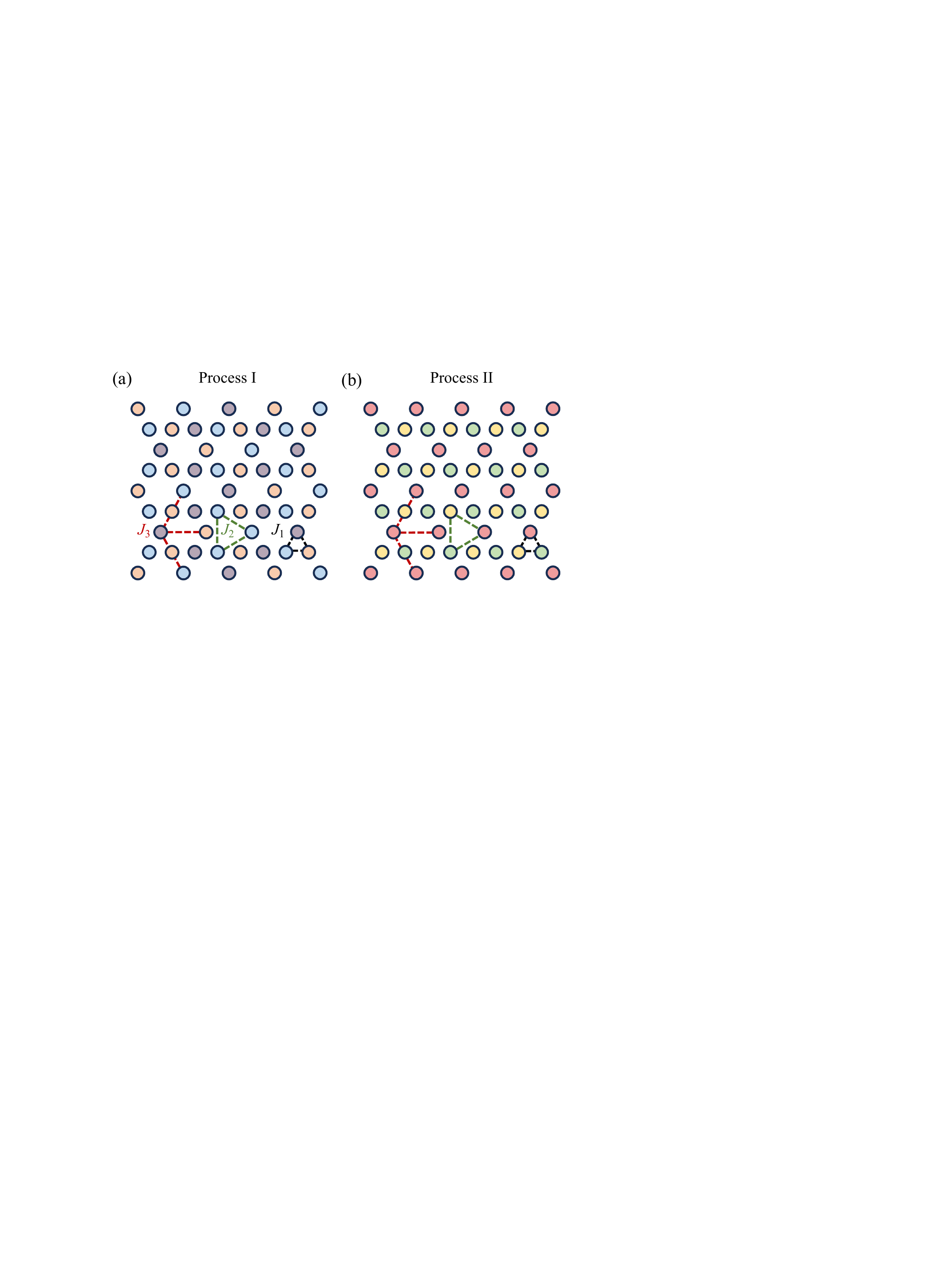}  
    \caption{Renormalization of  $ J_1 $-$ J_2 $-$ J_3 $ spin interactions on the kagome lattice via dual-frequency modulation.  
    Site-resolved periodic modulations (amplitude $\delta_q$, frequency $\omega_q$, $q=1,2$) imprint phase (color) patterns $\phi_{mq}$ to engineer $J_1$-$J_2$-$J_3$ couplings independently.  
    (a) Tuning $J_2/J_1$: modulation at $\omega_1$ assigns phases $\phi_{m1} = 0$ (orange), $2\pi/3$ (gray), $4\pi/3$ (blue). Nearest- ($J_1$) and third-neighbor ($J_3$) interactions between distinct colors are renormalized by $\mathcal{N}_1$, while the $J_2$ interactions of same-color pairs remain unmodified.  
    (b) Tuning $J_3/J_1$: modulation at $\omega_2$ applies phases $\phi_{n2} = 0$ (red), $2\pi/3$ (green), $4\pi/3$ (yellow). First- ($J_1$) and second-neighbor ($J_2$) couplings between differing colors acquire a factor of $\mathcal{N}_2 $, while preserving $J_3$ (shared-color third neighbors).  
    Combined, these protocols independently adjust the interaction ratios $J_2^{\text{eff}}/J_1^{\text{eff}}$ and $J_3^{\text{eff}}/J_1^{\text{eff}}$, enabling controlled exploration of frustrated regimes.}  
    \label{fig2}  
\end{figure}

\textit{Kagome-lattice Heisenberg model.}---A paradigmatic platform for highly frustrated quantum magnetism is the kagome-lattice Heisenberg antiferromagnet~\cite{Sachdev92,yan2011spin}. The phase diagram of this celebrated model---which is also believed to describe certain layered compounds such as herbertsmithite~\cite{PhysRevLett.98.107204,han2012fractionalized}---is remarkably intricate. Even with only NN ($J_1$) interactions, the system can host a time-reversal-invariant U(1) Dirac spin liquid (DSL) state~\cite{Iqbal-2021}. On adding $J_2$ and $J_3$ interactions, the phase diagram is known to harbor a second distinct QSL phase, namely a gapped chiral spin liquid (CSL)~\cite{Sheng-kagome-prb,he2014a,gong2014-kagome,Hu-2015,he2015,zhu2015a-kagome,zhu2019} that spontaneously breaks time-reversal symmetry, in addition to the DSL phase~\cite{He-2017}.
Here, we illustrate a dual-frequency modulation protocol to engineer $J_1$-$J_2$-$J_3$ spin models on the kagome lattice with independent control over their ratios. 
As illustrated in Fig.~\ref{fig2}, our approach includes two independent periodic modulations: 

\textit{Process I (frequency $\omega_1$)}: A site-resolved modulation $\delta_1 \sin(\omega_1 t + \phi_m)|nS\rangle\langle nS|$ assigns phases $\phi_m = 0$, $2\pi/3$, or $4\pi/3$ to lattice sites (orange, gray, and blue in Fig.~\ref{fig2}(a), respectively). Interactions on neighboring sites acquire a renormalization factor of $\mathcal{N}_1 = \mathcal{J}_0\left(\frac{2\delta_1}{\omega_1}\sin\frac{\pi}{3}\right)$, while  pairs with the same phase are unaffected. This selectively modifies $J_1$ and $J_3$ interactions, leaving $J_2$ unchanged.  

\textit{Process II (frequency $\omega_2$):} A complementary modulation $\delta_2 \sin(\omega_2 t + \phi_n)|nS\rangle\langle nS|$ (Fig.~\ref{fig2}(b)) assigns different phases to NN and 2NN sites, thus imprinting a factor of $\mathcal{N}_2 = \mathcal{J}_0\left(\frac{2\delta_2}{\omega_2}\sin\frac{\pi}{3}\right)$ on $J_1$ and $J_2$, while $J_3$ interactions between sites that share a common phase are unchanged.  

The combination of these two processes renormalizes the interaction coefficients as 
$
J_1^{\text{eff}} = J_1 \mathcal{N}_1\mathcal{N}_2,   
J_2^{\text{eff}} = J_2 \mathcal{N}_2, 
J_3^{\text{eff}} = J_3 \mathcal{N}_1,  
$
yielding the tunable ratios:
\begin{equation}  
\frac{J_2^{\text{eff}}}{J_1^{\text{eff}}} = \frac{1}{\mathcal{N}_1} \frac{J_2}{J_1}, \quad  
\frac{J_3^{\text{eff}}}{J_1^{\text{eff}}} = \frac{1}{\mathcal{N}_2} \frac{J_3}{J_1},  
\end{equation}  
where $\mathcal{N}_1$ and $\mathcal{N}_2$ are independently controlled by adjusting $\delta_1/\omega_1$ and $\delta_2/ \omega_2$. 
This enables suppression or enhancement of $J_2/J_1$ and $J_3/J_1$ across different parameter regimes as required. 
In the absence of Floquet driving, the ratio of couplings $J_1$:$J_2$:$J_3$ is set by the dipolar interactions to be $1\,$:$\,3^{-3/2}\,$:$\,2^{-3} \simeq 1\,$:$\, 0.192\,$:$\,0.125$, which restricts the tunability of the frustration. For instance, in the isotropic Heisenberg model, the CSL phase emerges only for $J_3 \gtrsim J_2$~\cite{Hu-2015}; even in the XY limit ($J^z_\mu$\,$=$\,$0$), the native dipolar system lies very close to the CSL--DSL phase boundary~\cite{he2015}.
Our protocol, however, enables access to interaction ratios unattainable in static kagome systems, such as $ J_2/J_1 \!\sim\! 0.2 $ and $ J_3/J_1 \!\sim\! 0.2 $, which positions us robustly within a stable CSL phase~\cite{gong2014-kagome,Sheng-kagome-prb,zhu2019}.
Besides, unlike materials with immutable spin couplings, this scheme provides \textit{in situ} control, allowing one to dynamically drive a quantum phase transition from a prepared ``simple" phase, such as magnetically ordered states, into regimes with nontrivial topological order~\cite{coarsening2024}. 
In addition, the system's inherent longer-range interactions---
including two symmetry-inequivalent $J_3$
and weaker higher-order (e.g., $J_1^{\text{eff}} / J_4^{\text{eff}} \approx 18.5$)
---make the dipolar model distinct from conventional $J_1$–$J_2$–$J_3$ models and lead to rich, exotic quantum phases in such frustrated spin systems~\cite{Yao2018,lugan2022,colbois2022}. While we primarily discuss the $J_1$-$J_2$-$J_3$ control,
flexible tuning of the distinct $J_3$ terms~\cite{supplementary},
along with residual $J_l$ terms for $l > 3$, will also be important when considering the precise many-body phases supported under such Floquet control.

\begin{figure*}
    \centering
    \includegraphics[width=\linewidth]{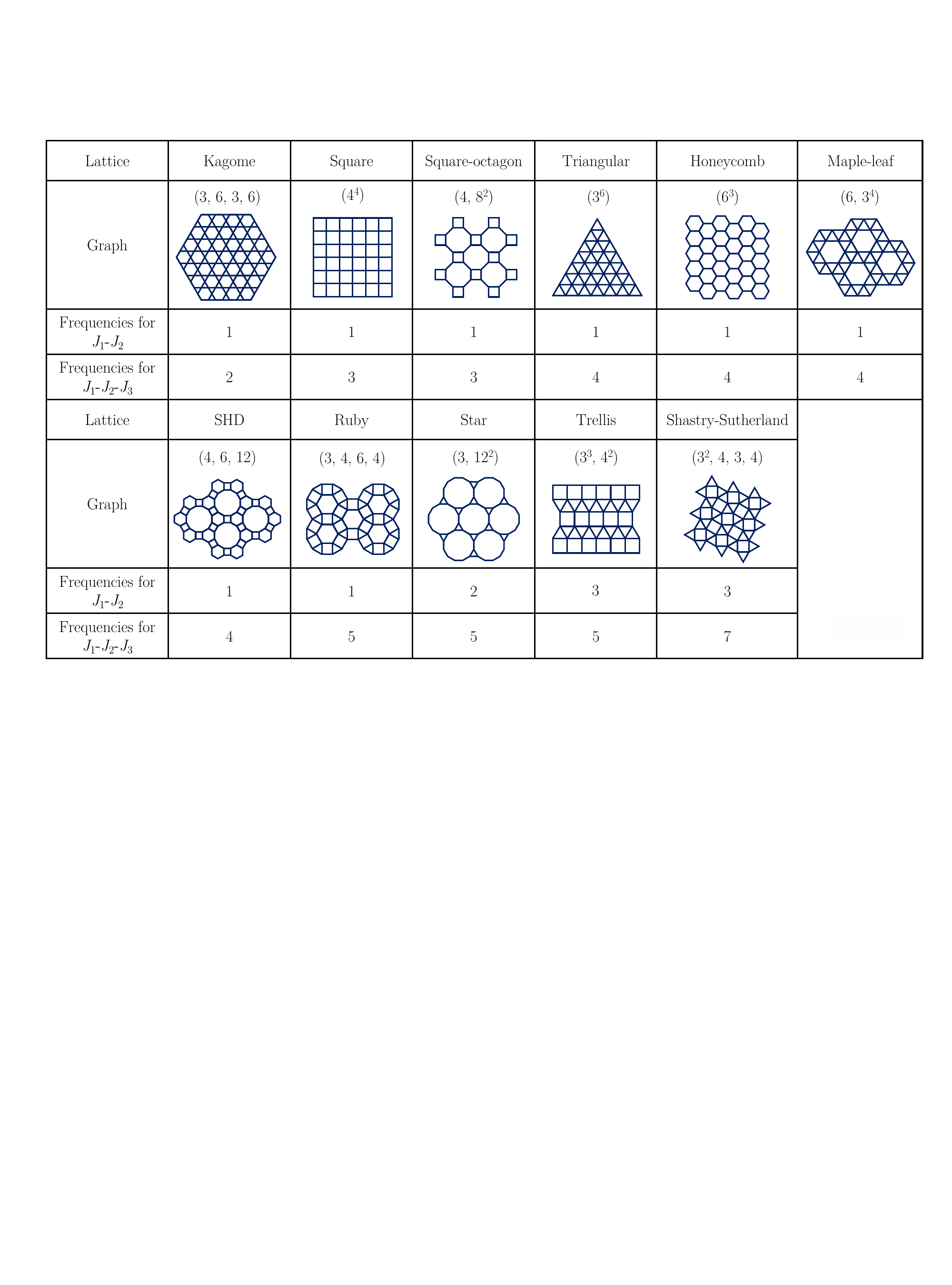}
    \caption{The required frequencies for engineering long-range interactions in 11 Archimedean lattices. 
  All lattices are vertex-transitive, with their mathematical descriptions (in brackets) provided by sequences of numbers $ n_i $, separated by commas (e.g., $ n_1,n_2,\cdots,n_r $), representing the number of vertices in the polygons surrounding each vertex of the lattice.
 Below each lattice graph, the number of modulated frequencies needed for controlling $ J_1 $-$ J_2 $ and $ J_1 $-$ J_2 $-$ J_3 $ interactions is listed.
    }
    \label{fig3}
\end{figure*}

\textit{Graphically designing spin interactions}.---
The renormalization mechanism for the kagome lattice 
suggests a broad and straightforward guiding principle for 
engineering spin-spin interactions beyond nearest neighbors: (i) identify the two-process color patterns for a given lattice structure; (ii) design the onsite periodic modulation strategy for each color pattern.
For step (i), a general method is to map the lattice structure onto a graph $ G = (V, E) $, where vertices $ v \in V $ represent lattice sites and edges $ e \in E $ encode the interaction types (i.e., NN, 2NN, or 3NN). 
Selectively modulating $ J_1 $ and $ J_2 $ while keeping $ J_3 $ invariant requires vertices connected by third-neighbor edges ($ e_{mn} \in J_3 $) to share the same color, while vertices linked by first- or second-neighbor edges ($ e_{mn} \in J_1, J_2 $) adopt distinct colors. This is a graph-coloring problem. The symmetry and connectivity of the graph guide such color filling, enabling us to design the two-process color patterns.

For step (ii), the key principle is to ensure uniform modulation interaction strength for atom pairs separated by the same distance.
For lattices with simple color patterns, such as three-color patterns in Fig.~\ref{fig2}(a), there are three types of connections (represented by red or black dashed lines), and the renormalization factors for atom pairs are given by Eq.~(\ref{eqj}).
In this case, uniform modulation can be simply achieved using a single frequency, by assigning the phases $ \phi_m = 0, 2\pi/3, 4\pi/3 $ to the three colors, respectively.
However, for more complex colorings (e.g., four-color patterns), a single frequency is insufficient to uniformly modulate all types of links. 
A systematic solution is to use multifrequency modulation, with details provided in the \textit{End Matter} and SM~\cite{supplementary}.

For the most commonly studied lattices---to wit, the kagome, square, triangular, and honeycomb---four-color modulation patterns (or fewer) suffice due to their innate symmetries. This opens the door to accessing a wide variety of correlated many-body phases with Rydberg simulators. For instance, the square-lattice Heisenberg antiferromagnet hosts Néel-ordered and valence bond solid ground states for varying $J_2/J_1$~\cite{chan2023-square,square-qian2024}---with a potential gapless $\mathbb{Z}_2$ QSL \cite{Hu-2013, ferrari2018spectral} sandwiched between them---while the addition of a nonzero $J_3/J_1$ stabilizes an extended region of the gapless QSL phase~\cite{square-liu2024}.
Similarly, the $J_1$-$J_2$ triangular-lattice model exhibits a 120$^\circ$ N\'eel-ordered antiferromagnetic ground state and a two-sublattice collinear magnetically ordered state~\cite{zhu2015b-tri}, which also competes with a QSL phase~\cite{gallegos2024} that is believed to be a U(1) DSL~\cite{Hu-2019}. With $J_1$-$J_2$-$J_3$ interactions, arguments for both a CSL with spinon Fermi surfaces~\cite{PhysRevB.100.241111} and a fully symmetric QSL~\cite{PhysRevB.107.L140411} have been proposed.
Finally, on the bipartite honeycomb lattice, while various magnetic phases with N\'eel, valence-bond, and collinear orders have been recognized~\cite{honeycomb-oitmma2011,honeycomb-reuther2011}, recent studies have suggested the emergence of a gapless nonmagnetic phase \cite{PhysRevB.96.104401} in the $J_1$-$J_2$ model, as well as  QSLs with more complex gauge groups in the easy-axis (Ising) regime~\cite{giudice2022trimer,kornjavca2023trimer}.

Going beyond these relatively simple lattices, even in  geometries requiring up to six-color patterns (with fifteen distinct connections), we show that only four frequencies are needed to achieve uniform modulation for one process~\cite{supplementary}. 
Figure~\ref{fig3} shows the frequency requirements for controlling $J_1$-$J_2$ (one process) and $J_1$-$J_2$-$J_3$ (two processes) interactions across all 11 Archimedean lattices~\cite{yu2015ising,farnell2018} based on the above approach. Besides the four considered earlier, this family of vertex-transitive geometries includes the square-octagon~\cite{kargarian2010topological}, maple-leaf~\cite{PhysRevB.108.L060406,gresista2023candidate,sonnenschein2024candidate,ghosh2024-maple,ghosh2022-maple}, square-hexagonal-dodecagonal (SHD)~\cite{tomczak1999ground}, ruby~\cite{ruby, maity2024}, star~\cite{jahromi2018spin}, trellis~\cite{miyahara1998quantum,yamaguchi2018magnetic}, and Shastry-Sutherland~\cite{yang2022quantum,viteritti2024,qian2024,SSDirac} lattices, which have all been predicted to host a panoply of interesting magnetic phases.
Finally, we emphasize that this Floquet renormalization of spin-spin interactions in a geometrically defined manner is broadly applicable, beyond spin-$1/2$ systems~\cite{Andriy} and beyond planar lattices~\cite{StuffHC} and ordered lattices more generally.

\textit{Experimental feasibility.}---To implement the proposed modulation scheme, the Floquet period should be much shorter than the Rydberg state lifetime. Experimentally, modulation periods $\sim$~50\,ns (frequencies $\omega_q\sim$~20\,MHz) are readily achievable with acousto-optic devices, well below typical Rydberg lifetimes ($10$--$100\,\mu$s)~\cite{saffman2010}.
The modulation amplitude $\delta_q$, set by the AC Stark shift, can be tuned over a broad range---from a few kHz to several tens of MHz---by varying the intensity and detuning of the addressing lasers~\cite{shift-sylvain}. This enables $\delta_q/\omega_q$ to vary continuously from 0 up to order unity, providing wide tunability of the effective interactions~\cite{supplementary}.
To suppress unwanted tunneling terms, the modulation frequency $\omega$ should satisfy $\Omega_{\text{mw}} \gg \omega \gg J$, where $\Omega_{\text{mw}}$ is the microwave driving strength and $J$ is the interaction strength. For single-frequency modulation (e.g., when tuning $J_1$-$J_2$ only), 
this condition is straightforward to meet, as $J \sim 1$\,MHz is typical in Rydberg systems~\cite{browaeys2016} and $\Omega_{\text{mw}} \sim 100$\,MHz is readily achievable in larger bias magnetic fields. 
For multiple frequency modulation (e.g., in controlling $J_1$-$J_2$-$J_3$), the frequency differences $|\omega_i - \omega_j|$ should also obey $\Omega_{\text{mw}} \gg |\omega_i - \omega_j| \gg J$, requiring careful frequency selection but remaining experimentally accessible nonetheless.  
In this regime of $J\approx 1$ MHz, a lattice spacing of about 10 $\mu$m and an addressing laser beam waist of roughly 1-2~$\mu$m enable precise spatial control with negligible crosstalk to neighboring sites.

In dipolar Rydberg arrays, the required optical modulation patterns can be achieved by simply combining several identical modulated patterns, offset spatially and delayed in their modulation phase. In the complementary platform of dipolar molecule arrays~\cite{Cheuk-Array,Bao-array,Ruttley2025}, where the scale of exchange interactions is roughly four to six orders of magnitude smaller than in dipolar Rydberg systems, equivalent and even more generic modulation patterns can be achieved simply through direct time-dependent shaping of a single laser field by, e.g., a fast spatial light modulator.

\textit{Conclusion.}---We present a versatile framework for realizing the most general $J_1 $-$ J_2 $-$ J_3$ Heisenberg model in geometrically engineered Rydberg lattices.
By leveraging the use of microwave-dressed states~\cite{Gorsh-mol} and
local laser fields to periodically shift the Rydberg energy levels~\cite{shift-sylvain},
one can realize an effective XXZ or Heisenberg model where transverse spin-exchange terms are precisely engineered, with an effective parameter range beyond that provided by nature.  
We analyze this approach on the kagome lattice, a geometrically frustrated system, showing how $ J_1 $, $ J_2 $, and $ J_3 $ interactions can be independently controlled. Then, we extend the framework to a broad class of lattice geometries using graph theory, mapping lattice structures to graphs with site-dependent modulations determined through graph coloring. 
Beyond modulating $J_1$, $J_2$, and $J_3$, the approach can be generalized to control spin interactions in more complex scenarios, such as modulating symmetry-inequivalent bonds, as demonstrated in the Supplementary Material (SM)~\cite{supplementary}.

In the context of utilizing Floquet engineering to access new physics~\cite{Weitenberg2021}, spin systems such as Rydberg arrays stand out as promising platforms. For one, Floquet techniques~\cite{Goldman14} can be used to engineer new effective Hamiltonian terms~\cite{Paola-18,Weid-floq,Ryd-RAT,Lukin-Floquet-1,scholl2022,nishad2023a,kunimi2024},
without the deleterious higher-band heating effects that plague shaken-lattice experiments~\cite{Reitter}.
Furthermore, 
the long-range nature of the interactions even allows for new physics to be accessed by simply changing the relative strengths of existing, competing Hamiltonian terms~\cite{Lee-Floquet-manage,Surace-Integr,Zhao2023}.
Beyond simply reducing the strength of certain interaction terms, Floquet renormalization can also change their sign, which, e.g., could aid the exploration of isotropic dipolar spin interactions in three dimensions.
Besides, such tunability may further be used for exploiting long-range interactions in quantum metrology~\cite{bornet2023a-dynamics,block2024-np,chu2023-prl-metrology,zwierz-prl2010-metrology} and nonequilibrium physics~\cite{tan2021a-nonequlibrium,block2022a-nonequlibrium,minato2022-nonequlibrium,kranzl2023a-nonequlibrium}.

\begin{acknowledgments}
We acknowledge helpful and enlightening discussions with Chenxi Huang and Yasir Iqbal, and with Pratyay Ghosh on the maple-leaf lattice.
This work (M.T. and B.G.) is supported by the AFOSR MURI program under agreement number FA9550-22-1-0339. R.S. is supported by the Princeton Quantum Initiative Fellowship.
\end{acknowledgments}

\textit{Data availability.}---The data that support the findings of
this Letter are openly available~\cite{mingshengtian2025rawdata}.

%

\newpage

\onecolumngrid

\vspace{1em}
\begin{center}
    {\large\textbf{End Matter}}
\end{center}

\begin{figure*}[h]
    \centering
    \includegraphics[width=15cm]{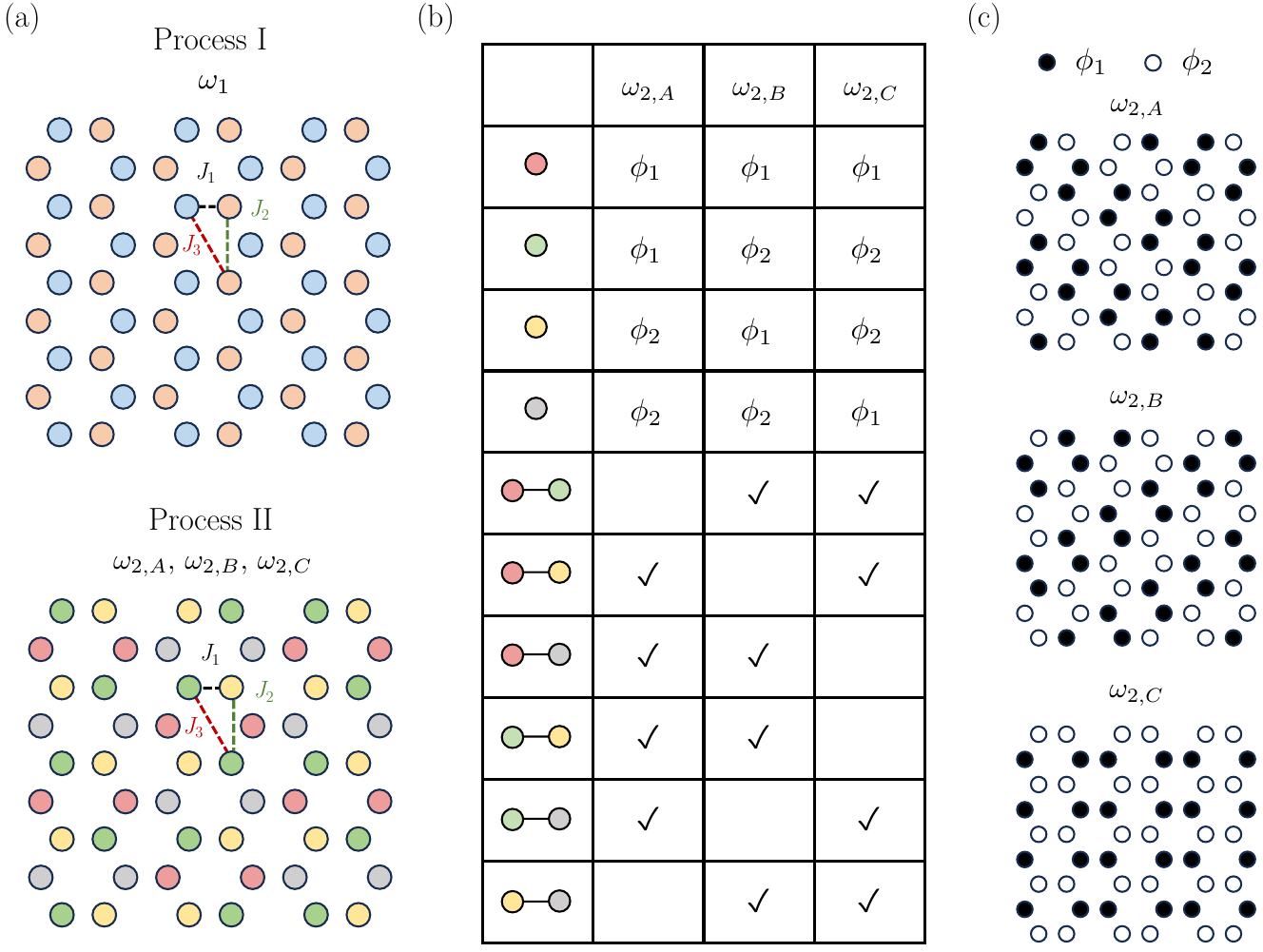}
    \caption{
    Strategies for controlling $J_1$-$J_2$-$J_3$ spin interactions on the honeycomb lattice via multifrequency modulation. 
    Each colored circle represents a distinct local modulation applied to the corresponding lattice site, expressed as 
    $\sum_{m,q} \delta_q \sin(\omega_q t + \phi_{mq})$, where $m$ indexes the lattice sites, $q$ indexes the modulation frequencies, and $\phi_{mq}$ denotes the modulation phase associated with site $m$ and frequency $q$.
    (a) The two processes for modulating $J_1$-$J_2$-$J_3$ interactions. In process I, a two-color pattern is designed to tune $J_2$/$J_1$: modulation at frequency $\omega_1$ assigns phases $\phi_1'$ for blue, and $\phi_2'$ for orange. $J_1$ and $J_3$ are renormalized by a factor $\mathcal{N}_1$ (Eq.~\ref{eqj}) while $J_2$ is unchanged.
    In process II, a four-color pattern is designed to tune $J_3/J_1$, modulated by three different frequencies.
    (b)~The designed modulation frequencies for the four-color lattice in process II. To ensure uniform modulation across each pair, we use three distinct frequencies $\omega_{2,A}$, $\omega_{2,B}$, and $\omega_{2,C}$. The corresponding modulation phases are listed in the table. The color pairings between modulated lattice sites are indicated, with a “$\checkmark$” denoting modulated interactions and a blank space indicating unmodulated ones.
    Each frequency modulates four connections, with the tiling ensuring uniform modulation across all six connections.
    (c)~The corresponding color pattern for three different frequencies in process II, wherein we decompose process II into three color patterns and use black and white to represent the phase difference for each frequency modulation.}
    \label{sfig2a}
\end{figure*}
\twocolumngrid
\textit{Appendix A: Multifrequency modulation strategy for complex color patterns.---}To achieve uniform modulation strength for atom pairs separated by the same distance, as required in step (ii) above for graphically designing spin interactions, a systematic solution is to apply multifrequency modulation, where each frequency targets a specific subset of connections. This combinatorial approach ensures uniform interaction strengths for all pairs at equal distances.

As a concrete example, we consider the four-color modulation pattern on a honeycomb lattice, illustrated in Fig.~\ref{sfig2a}. The full modulation consists of two processes. Process I is designed to tune the ratio $J_2/J_1$ using a two-color pattern, which can be implemented with a single frequency.
Modulation at frequency $\omega_1$ assigns phases $\phi_1'$ and $\phi_2'$ to the blue and orange sites, respectively, as shown in Fig.~\ref{sfig2a}(a). The interactions $J_1$ and $J_3$ are renormalized by a factor $\mathcal{N}_1$, given by Eq.~(\ref{eqj}).

Process II aims to tune $J_3/J_1$ using a four-color pattern, which requires multifrequency modulation. As illustrated in Fig.~\ref{sfig2a}(b), this configuration includes six distinct types of connections between the colored sites.
Frequency $\omega_{2,A}$ modulates all connections except red–green and yellow–gray; frequency $\omega_{2,B}$ excludes red–yellow and green–gray; and frequency $\omega_{2,C}$ excludes red–gray and green–yellow. Each connection is modulated by two of the three frequencies.
Assuming an identical renormalization factor $\mathcal{N}_0$ for each frequency (i.e., $\delta_q/\omega_q = \text{const.}$), the combined effective renormalization factor becomes $\mathcal{N}_0^2$, applied uniformly across all six connection types.
The resulting effective interaction strength is then given by:
\begin{equation}
   \mathcal{N}_{2} = \mathcal{J}_0^2\left(\frac{2\delta}{\omega} \sin\left(\frac{\phi_2 - \phi_1}{2}\right)\right),
\end{equation}
under the condition $\omega_q/\delta_q \equiv \omega/\delta$.
Following this principle, the required modulation frequencies for a given color pattern can be systematically determined. See the SM~\cite{supplementary} for a detailed discussion on more complex color patterns.

\end{document}


\title{Supplemental Material: Engineering frustrated Rydberg spin models by graphical Floquet modulation}    

\author{Mingsheng Tian}
\affiliation{Department of Physics, The Pennsylvania State University, University Park, Pennsylvania, 16802, USA}
\author{Rhine Samajdar}
\email{rhine\_samajdar@princeton.edu}
\affiliation{Department of Physics, Princeton University, Princeton, NJ 08544, USA}
\affiliation{Princeton Center for Theoretical Science, Princeton University, Princeton, NJ 08544, USA}
\author{Bryce Gadway}
\email{bgadway@psu.edu}
\affiliation{Department of Physics, The Pennsylvania State University, University Park, Pennsylvania, 16802, USA}

\maketitle
\tableofcontents

\section{Derivation of the effective Hamiltonian}

\subsection{Mapping to the XXZ model}
The dipole-dipole interaction between two Rydberg atoms placed at positions $\boldsymbol{r}_m$ and $\boldsymbol{r}_n$, with electric dipole operators $\hat{\boldsymbol{d}}_m$ and $\hat{\boldsymbol{d}}_n$, is expressed as:
\begin{equation}
    \hat{V}_{\mathrm{ddi}}=\frac{\hat{\boldsymbol{d}_m} \cdot \hat{\mathbf{d}_n}-3\left(\hat{\boldsymbol{d}_m} \cdot \boldsymbol{n}\right)\left(\hat{\boldsymbol{d}_n} \cdot \boldsymbol{n}\right)}{4 \pi \epsilon_0 R^3},
\end{equation}
with $R\equiv\left|\boldsymbol{r}_{\mathbf{2}}-\boldsymbol{r}_{\mathbf{1}}\right|$ being the distance between the atoms, and $\boldsymbol{n}\equiv\left(\boldsymbol{r}_{\mathbf{2}}-\boldsymbol{r}_{\mathbf{1}}\right) / R$ the unit vector defining the internuclear axis.
Here, we consider dipole-dipole interactions in a perpendicular magnetic field (which sets the quantization axis), where two atoms are placed in the $(x, y)$ plane and submitted to a static magnetic field $B_z$ along the $z$ direction. The unit vector defining the internuclear axis is $\boldsymbol{n}=(\cos \phi, \sin \phi, 0)$, with $\phi$ the angle between the internuclear axis and the $x$-axis. In this basis, the dipole-dipole interaction can be written using the components of the dipole operator $\hat{d}^x, \hat{d}^y$, and $\hat{d}^z$ :
\begin{equation}
    \hat{V}_{\mathrm{ddi}}=\frac{1}{4 \pi \epsilon_0 R^3}\left[\hat{d}_m^z \hat{d}_n^z+\frac{1}{2}\left(\hat{d}_m^{+} \hat{d}_n^{-}+\hat{d}_m^{-} \hat{d}_n^{+}\right)-\frac{3}{2}\left(\hat{d}_m^{+} \hat{d}_n^{+} e^{-2 i \phi}+\hat{d}_m^{-} \hat{d}_n^{-} e^{2 i \phi}\right)\right]
\end{equation}
with $\hat{d}_n^{ \pm}=\mp\left(\hat{d}_n^x \pm i \hat{d}_n^y\right) / \sqrt{2}$ for atom $n$. 
These terms can be classified as follows.
The first three terms couple states that conserve the total internal angular momentum of the two atoms. 
Both the $\hat{d}_m^z \hat{d}_n^z$ ($\Delta m_J = 0$ transitions, with $m_J$ the projection of the spin angular momentum) terms and the $\hat{d}_m^{+} \hat{d}_n^{-}+\hat{d}_m^{-} \hat{d}_n^{+}$ ($\Delta m_J =\pm 1$ transitions) terms lead to ``flip-flop'' interactions where pairs of atoms exchange their internal state, and will be referred to as the spin-exchange terms in the following.
The last two terms describe the spin-orbit coupling: those couple two-atom states with different internal angular momenta. The conservation of the total angular momentum requires that these terms carry a phase $e^{ \pm 2 i \phi}$. These latter terms---which do not conserve the internal angular momentum---are typically nonresonant in the presence of external fields (due to, e.g., Zeeman shifts), and can generally be ignored except in cases of fine-tuned resonances.

As we now focus on the simple spin-exchange terms of the dipole-dipole interaction, 
we restrict our analysis to three Rydberg levels labeled as $\ket{r_1}$, $\ket{r_2}$, and $\ket{r_3}$, i.e., $\ket{nS}$, $\ket{nP}$, and $\ket{n'S}$ in Fig.~1 of the main text, where $n$ and $n'$ are principal quantum numbers.  Each level also has a particular set of quantum numbers $\{J, m_J \}$, with the specific level choices determining the $C_3$ power-law coefficient of the interactions.
The spin-down state is defined as $\ket{\downarrow} = \ket{r_1}$.  
The spin-up state is chosen to be a microwave-field-dressed state of an Autler-Townes doublet, split by an energy of $\hbar \sqrt{4\Omega_{\text{mw}}^2 + D^2}$, which is generated by strongly driving the $\ket{r_2} \leftrightarrow \ket{r_3}$ transition using a microwave with Rabi frequency $\Omega_{\text{mw}}$ and detuning $D$.  
This spin-up state is chosen as one of the two resulting dressed eigenstates:  
$\ket{\uparrow} = \cos\theta \ket{r_3} - \sin\theta \ket{r_2}$ for the lower-energy branch  
or  
$\ket{\uparrow} = \sin\theta \ket{r_3} + \cos\theta \ket{r_2}$ for the higher-energy branch,  
where $\theta = \arctan(2\Omega_{\text{mw}} / D) / 2$. Physically, these microwave-dressed states of the bare Rydberg levels $\ket{r_2}$ and $\ket{r_3}$ can be understood as dipoles that oscillate in-phase (lower branch) and out-of-phase (higher branch) with the microwave driving field. Here, without loss of generality, we have implicitly assumed a microwave phase offset of $\pi$.

Choosing the lower-energy branch as the spin-up state, the relevant interaction strengths are defined to be
        $J_{mn}^{\pm}= \bra{\uparrow\downarrow} \hat{V}_{\text{ddi}} \ket{\downarrow\uparrow}_{mn} = \sin^2\theta \braket{r_1, r_2| \hat{V}_{\text{ddi}} | r_2, r_1}_{mn}$
        and 
$4J_{mn}^{z} = \braket{\uparrow\uparrow | \hat{V}_{\text{ddi}} | \uparrow\uparrow}_{mn} = 2\sin^2\theta \cos^2 \theta \braket{r_2, r_3 | \hat{V}_{\text{ddi}} | r_3, r_2}_{mn}$.
The parameter $\theta$, which depends on the ratio of the microwave Rabi frequency and the detuning, allows us to globally tune the relative strengths of the transverse and longitudinal interaction terms in this dipolar XXZ model.

With these definitions, the bare Hamiltonian can be written in the spin-1/2 basis as:
\begin{equation}\label{seq:eff1}
\begin{aligned}
        H^{}_{0}=&
    \sum_{m<n}J_{mn}^{\pm}\left(\sigma_m^{+} \sigma_n^{-}+\sigma_m^{-} \sigma_n^{+}\right)
    +
    4J_{mn}^z\left(\frac{\sigma_m^0 + \sigma_m^z}{2}\right)\left(\frac{\sigma_n^0 + \sigma_n^z}{2}\right),
    \\
    =&\sum_{m<n}
    J_{mn}^{\pm}\left(\sigma_m^{+} \sigma_n^{-}+\sigma_m^{-} \sigma_n^{+}\right)
    +
    J_{mn}^z\sigma_m^z\sigma_n^z
    +J_{mn}^z\left(\sigma_m^z\sigma_n^0+\sigma_m^0\sigma_n^z\right)
      +  J_{mn}^z\sigma_m^0\sigma_n^0,
\end{aligned}
\end{equation}
where $\sigma^{x, y, z, 0}$ are the usual Pauli and identity matrices acting on the spin-$1 / 2$ states $\ket{\downarrow}$ and $\ket{\uparrow}$, $\sigma_m^+=(\sigma_m^x+i \sigma_m^y)/2$, and $\sigma_m^-=(\sigma_m^x-i \sigma_m^y)/2$. Up to a constant term $\sim \sigma_m^0\sigma_n^0$, this is nothing but the Hamiltonian for the XXZ model if one neglects the terms $\sigma_m^z\sigma_n^0$ and $\sigma_m^0\sigma_n^z$, which may be compensated by single-particle local-field terms near the system boundary. Alternatively, the $\ket{\downarrow}$ state can similarly be formed from a microwave-dressed state to engineer the XXZ model exactly. The Heisenberg point in this model occurs for $J_{mn}^{\pm}/2 = J_{mn}^{z}$, or equivalently, 
$\bra{\uparrow\downarrow} \hat{V}_{\text{ddi}} \ket{\downarrow\uparrow}_{mn} = \bra{\uparrow\uparrow} \hat{V}_{\text{ddi}} \ket{\uparrow\uparrow}_{mn}/2$.

In using simple, globally defined dressed states and only applying spatial patterns of temporal modulation to the $\ket{nS}$ Rydberg level, we show in the following that one can control (through Floquet renormalization) the transverse spin-exchange terms but not the longitudinal Ising-like interactions. Qualitatively, this maps well onto scenarios described by hard-core bosons with tunable dispersion and off-site interactions. While we focus on this experimentally simple situation, we note that added control of the longitudinal spin-spin interactions can be enabled through local and temporal control of the dressing parameter $\theta$.

\subsection{Floquet engineering the effective Hamiltonian}\label{sec:effHam}
We start with the Schrödinger equation with a time-periodic potential $V(t)$:
\begin{equation}
i \partial_t \psi(t)  =H(t) \psi(t),\,\text{with}\,\,
H(t)  =H_0+V(t) \, \text{and} \,\, V(t+T)=V(t).
\end{equation}
By considering the unitary transformation
$    \phi(t)=\mathcal{U}(t) \boldsymbol{\psi}(t)=e^{i {K}(t)} \boldsymbol{\psi}(t)$,
the new state $\phi(t)$ satisfies the Schrödinger equation
\begin{equation}\label{seq5}
    \begin{aligned}
i \partial_t \phi(t) & =H_I \phi(t),
\\
H_I & =e^{i  {K}(t)}  {H}(t) e^{-i  {K}(t)}+i\left(\frac{\partial e^{i  {K}(t)}}{\partial t}\right) e^{-i  {K}(t)},
\end{aligned}
\end{equation}
where we have introduced the interaction Hamiltonian $ H_I$. 
Specifically, if one considers $K'(t)=V(t)$, the interaction Hamiltonian simplifies to the form of:
\begin{equation}\label{seq6}
    H_I=\mathcal{U}(t) H_0 \,\mathcal{U}^\dagger(t)
    =e^{i  {K}(t)}  {H}_0 e^{-i  {K}(t)}.
\end{equation}
In our case, $ V(t) $ represents the temporal-modulation term, while $ H_0 $ corresponds to
the dipolar exchange interactions that underlie both the flip-flop and Ising-like terms of the XXZ model. By engineering specific forms of $V(t)$, we can flexibly tune the dipolar interaction terms.  Here, we consider unbiased cases ($ \langle V(t) \rangle = 0 $), where the renormalized exchange terms are well described as simply being scaled by zeroth-order Bessel functions with tunable modulation indices~\cite{eckardt2005,lignier2007}.

Specifically, to achieve flexible control over these interactions in our model, a time-dependent site-specific energy modulation is added solely to the spin-down state (i.e., Rydberg state $\ket{r_1}\equiv \ket{nS}$), with 
\begin{equation}\label{seq7}
V(t) = 
\sum_{m,q} \delta^{}_q \sin(\omega^{}_q t + \phi_{mq}) 
\ket{\downarrow}^{}_m\bra{\downarrow}^{}_m
=
\sum_{m,q} \delta^{}_q \sin(\omega^{}_q t + \phi^{}_{mq}) 
\sigma_m^- \sigma_m^+,
\end{equation}
where $m$ indexes the lattice sites and $q$ represents distinct modulation frequencies.
As we describe below, the different frequencies are assumed to be incommensurate so as to treat them as effectively independent modulation processes (ignoring weak higher-order multiphoton processes).

As mentioned in the main text, the local energy shifts $\delta$ can be applied as in Ref.~\onlinecite{shift-sylvain}, through an AC Stark shift induced by off-resonantly coupling to another internal level with focused laser beams. 
The nominal choice is to couple off-resonantly to the intermediate state (labeled as $\ket{p}$ in Fig.~1 of the main text) that also serves as part of a two-photon excitation scheme. In this case, using tightly focused lasers with micron-scale waists can realistically achieve resonant Rabi coupling strengths $\Omega_\textrm{addr}$ of up to hundreds of MHz, and one can detune from the $\ket{p}$ state on the scale of hundreds to thousands of MHz to ensure an off-resonance condition, $\Delta \gg \Omega_\textrm{addr}$.
Here, the induced shift will be $\delta \approx \Omega_\textrm{addr}^2 / \Delta$, on the scale of several tens of MHz. On intermediate timescales and for large detunings, one can ignore Rydberg de-excitations induced by this coupling light, which scale as $\gamma_\textrm{ind} \approx \Gamma_p ({\Omega_\textrm{addr}}/{\Delta})^2$, with $\Gamma_p$ the decay rate of the intermediate state $\ket{p}$ (being roughly $2\pi \times 1$~MHz for rubidium and potassium). For large detunings, this induced decay will remain smaller than the natural decay rate of the Rydberg levels, and thus, is not a major added limitation to the timescales for coherent evolution. However, when working with a fixed number of $\ket{\uparrow}$ excitations, one also has the ability to perform a judicious choice of measurement so as to post-select for cases in which no induced decay of the $\ket{\downarrow}$ atoms has occurred.

{By modulating the Rabi frequency $\Omega_{\text{addr}}(t)$ (via the laser intensity)  using fast acousto-optic or electro-optic (amplitude) modulators (AOMs or EOMs), one can generate the desired time-periodic AC Stark shifts, thereby realizing Floquet modulation. Each modulation corresponds to a distinct spatial pattern (color), and the required global optical modulation patterns can be achieved by simply combining several different modulation patterns. As this scheme is based on optical fields, the phase control can be implemented with high spatial and temporal precision, with negligible associated errors. To note, the phase of the modulation is not controlled by the actual phase of the laser fields, but rather by the phase of the laser intensity modulation (with modulation at a frequency $\omega_{q}$). Here, we note several technical considerations. In one possible implementation, the separate phases $\phi_{mq}$ would be individually controlled via separately driven AOMs or EOMs. These relative phases can be calibrated through the atomic response, and the phase relationships would be quite stable for the expected modulation frequencies (on the few-to-tens of MHz scale), insensitive to incidental changes in the optical path lengths (considering the optical paths being separately fiber-coupled to the experimental setup, for example). The magnitude of the modulation depth, $\delta_{mq}$, would be the most likely quantity to suffer from instabilities, as it depends on the overlap with the tweezer-prepared atoms and would likely be subject to some degree of non-common-mode alignment noise or drift. This issue could be mitigated through the use of flat-top Stark-shifting beams, which would allow for the individual addressing of atoms while permitting some tolerance to the exact co-alignment with the tweezers used for loading the atom array.}

{Having discussed these technical considerations, we now return to the Floquet-engineering scheme} based on Eqs.~(\ref{seq5}) and (\ref{seq6}), {considering} a rotating frame with
\begin{equation}
    K(t)=\sum_{m,q} -\frac{\delta^{}_q}{\omega^{}_q} \cos(\omega^{}_q t + \phi^{}_{mq}) \,\sigma_m^- \sigma_m^+.
\end{equation}
Then, the interaction Hamiltonian takes the form
\begin{equation}
\begin{aligned}
        H^{}_I=&
        e^{-i\sum_{n',q} \frac{\delta_q}{\omega_q} \cos(\omega^{}_q t + \phi^{}_{n'q}) \sigma_{n'}^- \sigma_{n'}^+}
       \left( 
       \sum_{m<n}J_{mn}^{\pm}\left(\sigma_m^{+} \sigma_n^{-}+\sigma_m^{-} \sigma_n^{+}\right)
       +
    J_{mn}^z\sigma_m^z\sigma_n^z
    \right)
        e^{i\sum_{m',q} \frac{\delta_q}{\omega_q} \cos(\omega^{}_q t + \phi^{}_{m'q}) \sigma_{m'}^- \sigma_{m'}^+}
        \\
        =&
        \left(\sum_{m<n}  J_{mn}^{\pm}
        e^{-i\left(\sum_q \frac{\delta_q}{\omega_q}\cos (\omega_q t+\phi_{nq})\right)}
        e^{i\left(\sum_q\frac{\delta_q}{\omega_q}\cos (\omega_q t+\phi_{mq})\right)}
        \sigma_m^+ \sigma_n^- +\mathrm{h.c.}  \right)
        +J_{mn}^z\sigma_m^z\sigma_n^z,
\end{aligned}
\end{equation}
so the transverse spin-spin interactions are renormalized by a factor 
$\mathcal{N}_{mn}$
with
\begin{equation}
\mathcal{N}_{mn}= 
        \exp\left[i\left(\sum_q \frac{\delta_q}{\omega_q}
        \left(
        \cos (\omega_q t+\phi_{mq})
        -\cos (\omega_q t+\phi_{nq}\right)
        \right)\right].
\end{equation}
Using the identity $\cos{A}-\cos{B}=-2\sin(\frac{A-B}{2})\sin({\frac{A+B}{2}})$, and the Jacobi-Anger identity
$$
e^{iz\sin\theta}=\sum_{\beta=-\infty}^{\infty} \mathcal{J}_\beta(z)e^{i\beta\theta},
$$
we have
\begin{equation}\label{seq13}
\begin{aligned}
        \mathcal{N}_{mn}&=
            \exp \left(i\sum_q \frac{2\delta_q}{\omega_q}
     \sin \left(\omega_q t+\frac{\phi_{mq}+\phi_{nq}}{2}\right)
        \sin \left(\frac{\phi_{nq}-\phi_{mq}}{2}\right)
        \right)
        \\
        &=\prod_q\left(
        \sum_\beta \mathcal{J}_\beta\left(\frac{2\delta_q}{\omega_q}\sin (\frac{\phi_{nq}-\phi_{mq}}{2})\right)e^{i\beta \left(\omega_q t+\frac{\phi_{mq}+\phi_{nq}}{2}\right)}
        \right).
\end{aligned}
\end{equation}
Under the conditions that the modulation frequencies are incommensurate and that the rotating-wave approximation (RWA) applies, the higher-order Bessel functions can be neglected, allowing the expression above to be represented by a zeroth-order Bessel function~\cite{eckardt2005,lignier2007,aidelsburger2013}, as follows:
\begin{equation}\label{seqNmn}
    \mathcal{N}_{mn}
    =\prod_q
         \mathcal{J}_0\left(\frac{2\delta_q}{\omega_q}\sin \left(\frac{\phi_{nq}-\phi_{mq}}{2}\right)\right).
\end{equation}
This form suggests how the application of time-dependent modulation [Eq.~(\ref{seqNmn})] provides the means to control the transverse spin-spin interactions.
We note that, formally, the use of incommensurate driving frequencies leads to a breakdown of the RWA, as one can always find weak higher-order, multiphoton processes that approximately conserve energy. However, the above form describing renormalized interactions provides a faithful representation on the intermediate timescales relevant to experiments with dipolar Rydberg atoms, which we have confirmed with numerical simulations (see Fig.~\ref{sfig3}). While Floquet heating would be expected to occur on long timescales (due to the breakdown of the RWA) without mechanisms for entropy removal, such timescales are largely already precluded due to the intrinsic decay of the excited Rydberg levels.

In the next section, we discuss how time-dependent modulation offers great flexibility to engineer the dipolar exchange interactions for various lattice geometries.

\begin{figure*}
    \centering
    \includegraphics[width=18cm]{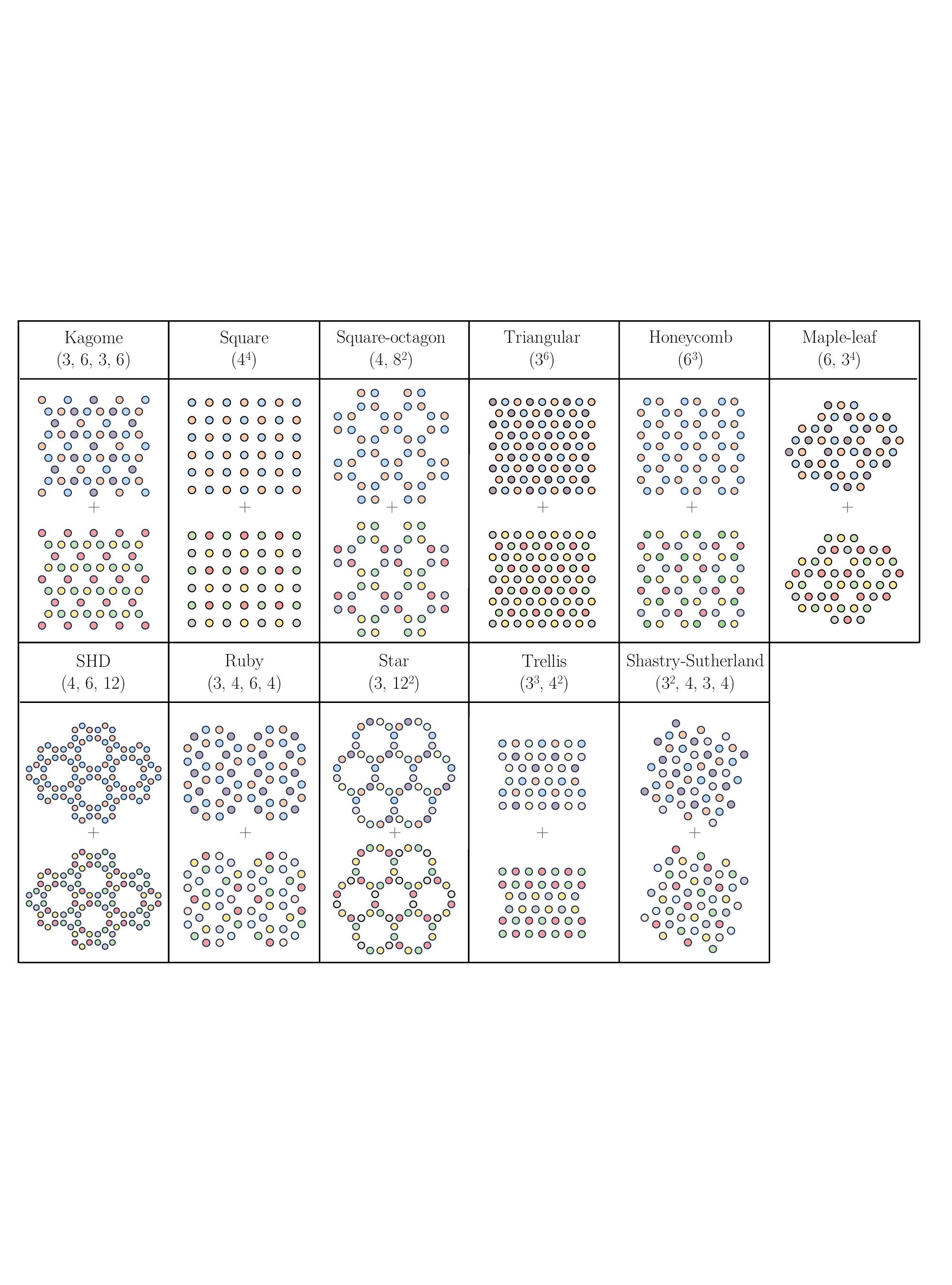}
    \caption{Color patterns used to engineer $J_1$-$J_2$-$J_3$ couplings in 11 Archimedean lattices. Each color denotes a distinct onsite modulation. 
    The first set of color patterns, for each lattice, are used for $J_1$-$J_2$ engineering, while the combination with the second pattern enables $J_1$-$J_2$-$J_3$ engineering. The mathematical description (in brackets) given by numbers $n_i$ separated by dots (i.e., $n_1,n_2,\cdots, n_r$) corresponds to the number of vertices of the polygons arranged around a vertex for each lattice.
    }
    \label{sfig1}
\end{figure*}

\begin{figure*}
    \centering
    \includegraphics[width=18cm]{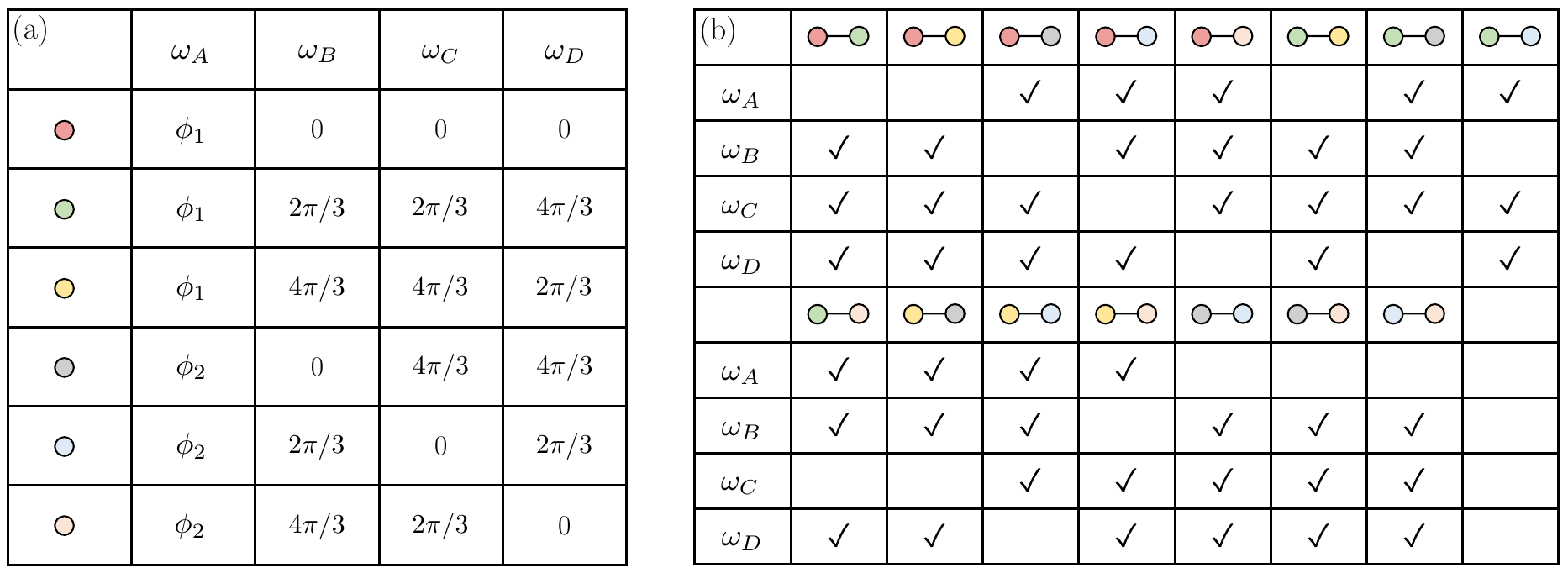}
    \caption{Protocols for uniformly modulating 15 pairs in the case of six-color patterns. Four distinct frequencies are used to ensure uniform modulation across all 15 unique connections. 
}
    \label{sfig2b}
\end{figure*}

\begin{figure*}
    \centering
    \includegraphics[width=18cm]{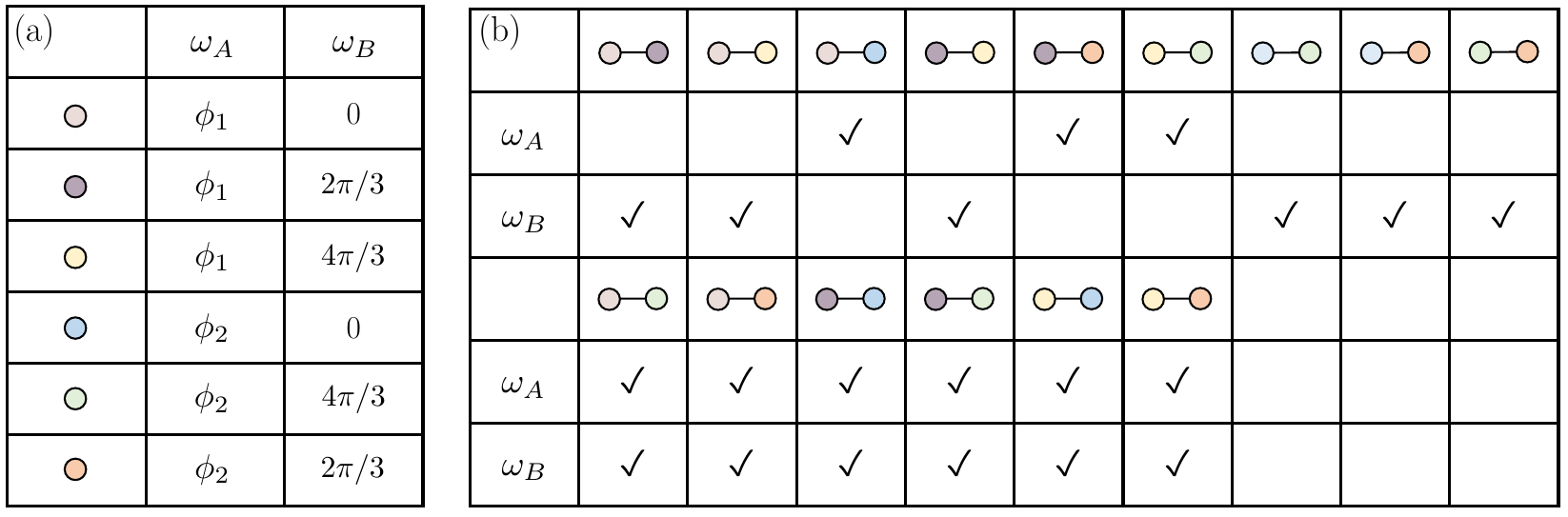}
    \caption{Protocols for nonuniformly modulating 15 pairs in the case of six-color patterns, where we uniformly modulate 9 pairs and 6 pairs by two different frequencies. This can be used for the star and trellis lattices, where 9 pairs can be designed for modulating $J_1$, while the remaining 6 are for $J_2$.}
    \label{sfig2c}
\end{figure*}

\section{Graphically engineering Hamiltonians for 11 Archimedean lattices}

In this section, we present detailed protocols for engineering the spin interactions for the 11 Archimedean lattices. The symmetries and graph properties of the lattices guide us in assigning color patterns, where each color corresponds to a unique local modulation. By choosing appropriate color arrangements, we can systematically control the ratios $J_2/J_1$ and $J_3/J_1$.

Figure~\ref{sfig1} shows the color patterns for each lattice. Usually, the first pattern focuses on $J_2/J_1$ engineering and can be achieved by coloring all next-nearest-neighbor sites with the same color, while ensuring nearest neighbors differ in color. The second pattern is used for $J_3/J_1$ engineering. In this case, the third-nearest neighbors share the same color, while nearest and next-nearest neighbors each have distinct colors. This strategy works for most Archimedean lattices, with the exception of some special structures; for instance, on the star lattice---due to its triangular-dodecagon structure---achieving identical coloring for all second-order neighbors is not feasible. 
Instead, we assign different colors to second-order neighbors while ensuring $J_1$ and $J_2$ remain distinguishable (a detailed method for handling this case is provided below).

To guarantee uniform modulation of $J_1$-$J_2$-$J_3$, it is essential to ensure that connected lattice sites that are the same distance apart experience identical modulation. According to Eq.~(\ref{seqNmn}), for two sites $m$ and $n$ with modulations $\delta \sin(\omega t+\phi_m)$ and $\delta \sin(\omega t+\phi_n)$, the effective modulation factor is given by
\begin{equation}\label{seq15}
   \mathcal{N}_{mn}= \mathcal{J}_0\!\left(\frac{2\delta}{\omega}\sin\!\left(
    \frac{\phi_{n}-\phi_{m}}{2}
    \right)
    \right).
\end{equation}
For two-color patterns, simply assigning distinct phases to the two colors is sufficient. For three-color patterns, evenly spacing the phases, such as $\phi_A = 0$, $\phi_B = 2\pi/3$, and $\phi_C = 4\pi/3$, ensures uniform modulation between all pairs of colors.
For lattices with more color patterns, such as four-color patterns with six distinct connections, a single modulation frequency is not enough. 
To address this case, we consider multifrequency modulation (as discussed in the \textit{End Matter}). Each of the frequencies modulates a subset of the connections, ensuring uniform modulation across all connections.

\begin{figure}[tb]
  \centering
  \includegraphics[width=12cm]{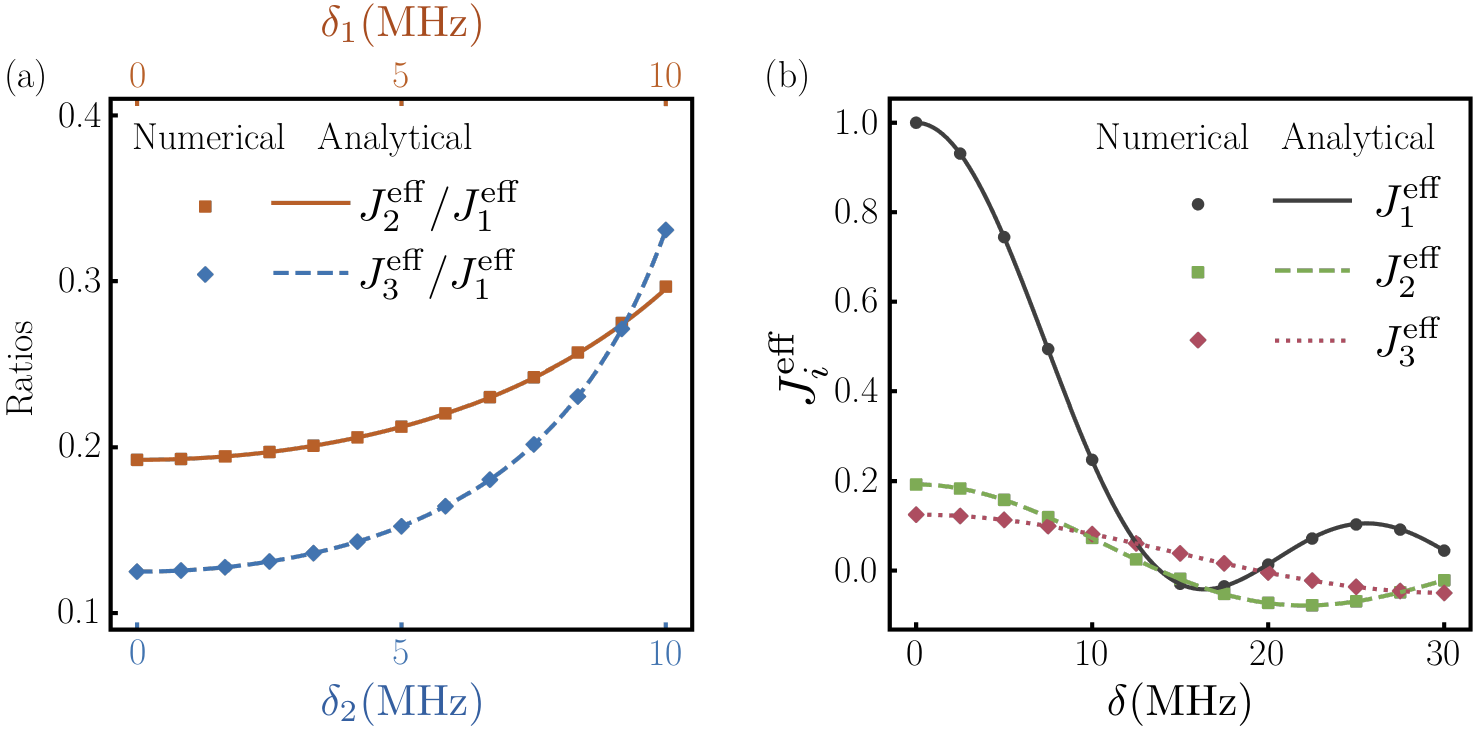}
  \caption{
Analytical and numerical results for the effective long-range interaction strengths in a driven kagome lattice using a two-frequency scheme.
  (a) Ratio of effective coupling strengths versus independent on-site energy shifts $\delta_1$ and $\delta_2$, with modulation frequencies $(\omega_1, \omega_2) = (14.14\ \mathrm{MHz}, 10\ \mathrm{MHz})$. Solid curves show the analytically calculated coupling strength with a zeroth-order Bessel function approximation (Eq.~(\ref{seqNmn})) and data points denote numerically calculated coupling strengths for a modulation duration of $T = 25\ \mu\mathrm{s}$ without such approximation (Eq.~(\ref{seq13})).   
  (b) Effective couplings $J_1$, $J_2$, and $J_3$ as functions of the onsite energy shift $\delta$ on a kagome lattice under modulation with frequencies $(\omega_1, \omega_2)$ and $\delta_1=\delta_2=\delta$. The remaining parameters are same as those in (a).}
  \label{sfig3}
\end{figure}

\begin{figure*}
    \centering
    \includegraphics[width=0.3\linewidth]{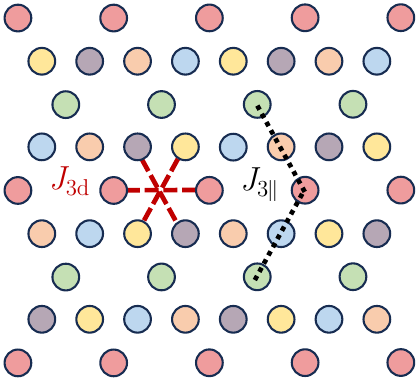}
    \caption{Six-color pattern for tuning the relative strengths of the two types of $J_3$ interactions on the kagome lattice, namely $J_{3d}$ (dashed red) and $J_{3\parallel}$ (dotted black). Atom pairs connected by $J_{3d}$ share the same color with no renormalization, whereas atom pairs connected by $J_{3\parallel}$ are assigned different colors, leading to a renormalized effective strength.}
    \label{sfig5}
\end{figure*}

\begin{figure*}
    \centering
    \includegraphics[width=0.75\linewidth]{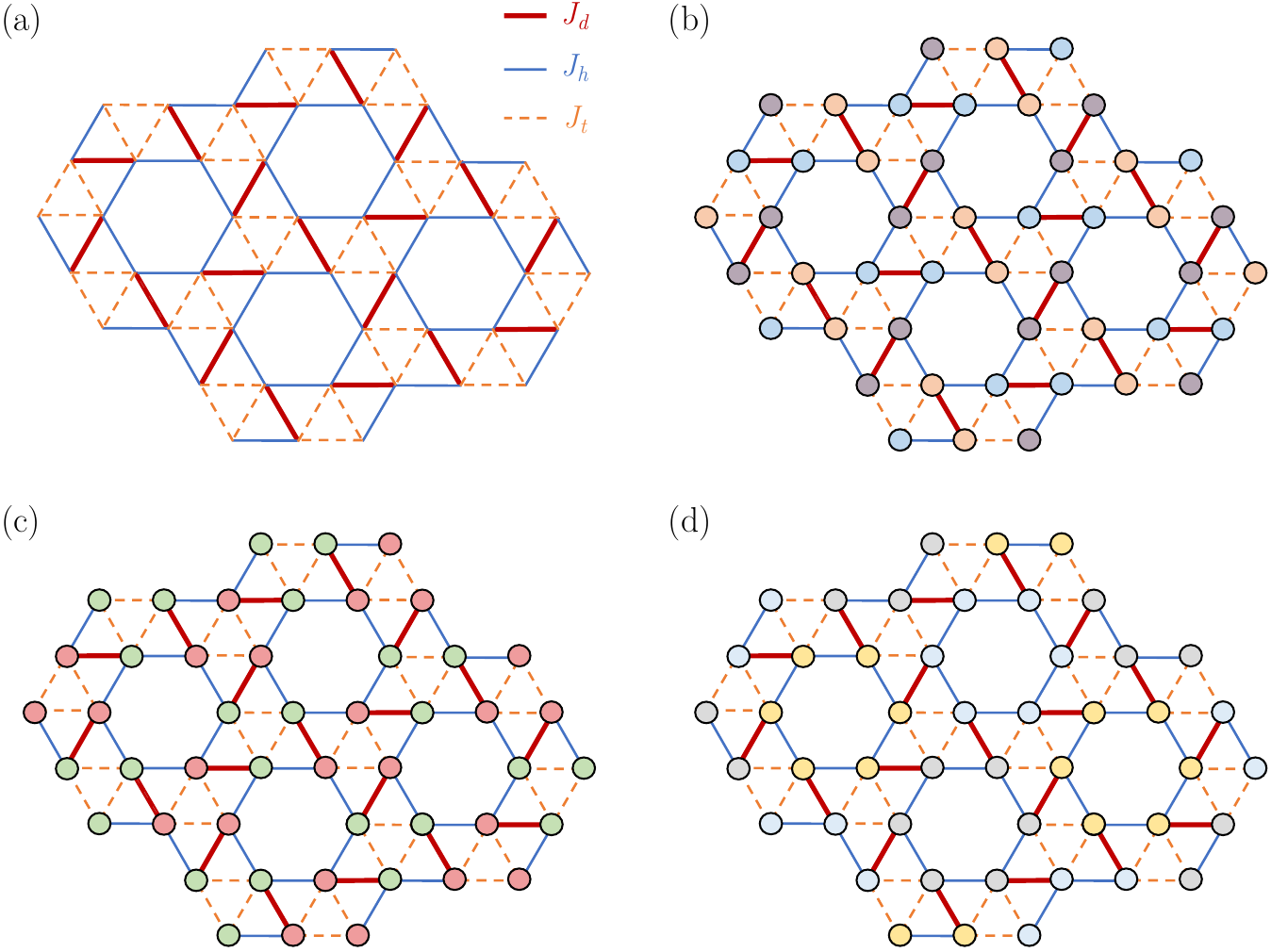}
    \caption{Protocols for controlling symmetry-inequivalent bonds on the maple-leaf lattice. Three different bonds---thick red ($J_d$), solid blue ($J_h$), and dashed orange ($J_t$)---are labeled in (a).  (b)~Designed three color-pattern to realize $J_d>J_t=J_h$ by renormalizing $J_t$ and $J_h$ with a factor $J_0(\sqrt{3}\delta/\omega)$. This control can lead to an exact dimer singlet ground state when $J_d$ is significantly larger than $J_t=J_h$~\cite{ghosh2022-maple}. (c,d) Complementary patterns selectively renormalize ($J_d$, $J_h$) and ($J_d$, $J_t$) respectively.
Combining any two of the three patterns enables flexible control of the $J_d$–$J_h$–$J_t$ interactions.
    }
    \label{sfig4}
\end{figure*}

As a detailed example, we show how to modulate the $J_1$-$J_2$-$J_3$ spin interactions on a honeycomb lattice by such a multifrequency process in the \textit{End Matter} of the main text.
When considering six-color patterns with up to 15 distinct connections, four frequencies are sufficient to achieve uniform modulation. Figure~\ref{sfig2b} outlines the phase arrangements and modulation strategies for these cases. Notably, not all four- or six-color patterns demand the maximum number of frequencies. The required frequencies depend on the number of distinct connections involved in interactions with different ranges. For example, in the four-color pattern of the square lattice shown in Fig.~\ref{sfig1}, $J_1$ involves only four connections, while $J_2$ involves two. In this case, two frequencies ($\omega_{2,A}$ and $\omega_{2,B}$ in Fig.~4 of main text) are sufficient, where $J_1$ is modulated once and $J_2$ is modulated twice. Similarly, in six-color patterns, such as those for the star and trellis lattices, two modulation frequencies can achieve uniform modulation across nine connections for $J_1$ and six for $J_2$, as shown in Fig.~\ref{sfig2c}.

Using these principles, we can determine the necessary modulation frequencies and design the corresponding color patterns with the two processes above; the spin interactions are renormalized as:
$
J_1^{\text{eff}} = J_1 \mathcal{N}_1\mathcal{N}_2; \quad
J_2^{\text{eff}} = J_2 \mathcal{N}_2;  \quad
J_3^{\text{eff}} = J_3 \mathcal{N}_1.
$
By independently adjusting $(\delta_i, \omega_i)$ in Eq.~(\ref{seqNmn}), one can control $\mathcal{N}_1$ and $\mathcal{N}_2$ to enhance or suppress specific interactions as shown in Fig.~\ref{sfig3}. In this example, we focus on the comparatively simple kagome lattice which requires only two-frequency modulation. These simulations confirm that the analytical expressions for the effective renormalized Hamiltonian are in good agreement with the numerical simulations of the system's dynamics on experimentally relevant timescales. Simulations of lattices requiring five-frequency modulation (not shown) similarly confirm the validity of the expressions for the renormalized interaction on such timescales.

Notably, our approach is not limited to modulating $J_1$, $J_2$, and $J_3$ interactions---it can be generalized to control spin interactions in broader scenarios, such as modulation of symmetry-inequivalent bonds with the same distance.  
As an example, we illustrate the color pattern in Fig.~\ref{sfig5} to showcase the control over two types of $J_3$ interactions on the kagome lattice. 
Combining with the color pattern in Fig.~2 of the main text, this enables us to control the four symmetry-inequivalent interactions $J_1$-$J_2$-$J_{3d}$-$J_{3||}$ on the kagome lattice. 
We note that achieving independent control over the two $J_3$ interactions motivates the exploration of additional spin models discussed in the condensed-matter literature~\cite{gong2014-kagome,lugan2022,colbois2022}, which may host a wider variety of intriguing quantum phases.
Another paradigmatic example is the maple-leaf [Fig.~\ref{sfig4}(a)], which has three distinct nearest-neighbor (NN) interactions: $J_d$ (thick red), $J_h$ (blue), and $J_t$ (dotted orange). When $J_d \gg J_t = J_h$, the model exhibits an exact dimer-singlet ground state~\cite{ghosh2022-maple}. This case can be achieved by using a single-frequency Floquet drive with a three-color phase pattern [Fig.~\ref{sfig4}(b)], with $2\pi/3$ phase difference.
Complementarily, alternative color patterns in Figs.~\ref{sfig4}(c,d) enable different control over the NN interactions. By combining two of the three patterns with two incommensurate frequencies, we can independently tune $J_d$, $J_h$, and $J_t$. 
This approach offers a way to study the rich phase transitions and exotic physics on the maple-leaf lattice, ranging from magnetic order to disordered phases~\cite{ghosh2024-maple}.


%